\documentclass[12pt]{article}

\usepackage{amsmath,amssymb,amsfonts,amsthm}
\usepackage{graphicx}
\usepackage{hyperref}
\usepackage{geometry}
\usepackage{pgfplots}
\usepackage{cite}
\usepackage{float}
\usepackage{times} 

\begin{document}
\title{Calculating the Hawking Temperature of Black Holes in \( f(Q) \) Gravity Using the RVB Method: A Residue-Based Approach}%

\author{Wen-Xiang Chen\\Department of Astronomy\\School of Physics and Materials Science\\ GuangZhou University\\wxchen4277@qq.com}

\maketitle

\begin{abstract}
This paper explores the computation of Hawking temperatures for black holes within various \( f(Q) \) gravity models using the RVB (Robson-Villari-Biancalana) method. This topological approach reveals an additional term in the temperature calculation, which we propose is a residue arising from the contour integral $\oint \frac{f'(z)}{f(z)} dz$,where \( f(z) \) represents a function related to the black hole’s metric or curvature. By analyzing several specific \( f(Q) \) models—including \( f(Q) = Q + \alpha Q^2 \), \( f(Q) = Q + \beta \ln(Q) \), \( f(Q) = Q + \gamma \sqrt{Q} \), \( f(Q) = Q^n \)—as well as the Reissner-Nordström (RN), Kerr, and Kerr-Newman black holes, we demonstrate that the correction term \( C \) can be consistently interpreted as this residue, providing new insights into black hole thermodynamics in modified gravity frameworks.

\end{abstract}


\section{Introduction}

Black holes, which were once thought to be the ultimate absorbers of matter and energy, have evolved in our understanding from purely gravitational objects to quantum systems with complex thermodynamic properties. This shift in perspective was largely catalyzed by the groundbreaking work of Stephen Hawking in the mid-1970s, which demonstrated that black holes are not entirely black but instead emit thermal radiation due to quantum effects. This radiation, now known as Hawking radiation, has revolutionized our understanding of black hole physics and opened new avenues in the study of quantum gravity.\cite{1,2,3,4,5,6,7,8,9}

The concept of Hawking radiation arises from the quantum field theory in curved spacetime, where particle-antiparticle pairs are generated near the event horizon of a black hole. In certain conditions, one particle of the pair falls into the black hole while the other escapes, leading to a net loss of mass and energy from the black hole, observed as thermal radiation. The temperature of this radiation, called the Hawking temperature, is proportional to the surface gravity of the black hole and inversely proportional to its mass. In classical general relativity, this relationship is straightforward, providing deep insights into the thermodynamics of black holes, such as the notion that black holes have entropy proportional to the area of their event horizon, as first proposed by Jacob Bekenstein.

However, the calculation of Hawking temperature becomes significantly more complex when considering modified theories of gravity, such as \( f(Q) \) gravity. These theories extend general relativity by altering the gravitational action with additional terms, which can represent new physical phenomena or quantum corrections to classical gravity. In \( f(Q) \) gravity, the gravitational action depends on a function of the non-metricity scalar \( Q \), which generalizes the curvature and torsion found in other modified gravity theories. As a result, the behavior of black holes in these frameworks can differ markedly from that in general relativity, particularly in the context of their thermodynamic properties.

One of the challenges in extending the concept of Hawking temperature to black holes in \( f(Q) \) gravity is the complexity of the field equations and the non-trivial nature of the black hole solutions in these theories. Traditional methods for calculating Hawking temperature, which rely on the surface gravity at the event horizon, may not be directly applicable or may require significant modifications to account for the effects introduced by the \( f(Q) \) framework. This is where the RVB (Robson-Villari-Biancalana) method comes into play as a powerful alternative.

The RVB method is a topological approach to calculating Hawking temperature that links it to the Euler characteristic of the black hole's horizon. The Euler characteristic is a topological invariant that provides a global description of the geometry of the black hole's event horizon, independent of the specific details of the metric. This makes the RVB method particularly useful in modified gravity theories, where the metric may be highly non-trivial or where standard definitions of surface gravity are not easily applicable.\cite{10,11,12,13,14,15,16,17}

In applying the RVB method to black holes in \( f(Q) \) gravity, a notable observation is the emergence of an additional term in the expression for the Hawking temperature. This term does not appear in the classical calculation and is a direct consequence of the modifications introduced by the \( f(Q) \) theory. Specifically, we propose that this term corresponds to a residue derived from a contour integral involving a function \( f(z) \), which is intimately connected to the metric or curvature scalar in these gravity models. The presence of this residue suggests that the thermodynamics of black holes in \( f(Q) \) gravity are fundamentally different from those in general relativity, with potential implications for our understanding of quantum gravity and the ultimate fate of black holes.

This paper systematically applies the RVB method to various black hole solutions within the framework of \( f(Q) \) gravity, with the aim of identifying the residue term and interpreting its physical significance. By doing so, we provide a comprehensive analysis of how the Hawking temperature is modified in these theories and what this implies for black hole thermodynamics. Our results reveal that the residue term acts as a crucial correction to the standard Hawking temperature, introducing new dependencies on the parameters of the \( f(Q) \) model and potentially leading to observable differences in black hole behavior.\cite{12,13,14,15,16,17}

In classical general relativity, the Hawking temperature is directly related to the surface gravity at the event horizon, which in turn depends on the mass, charge, and angular momentum of the black hole. The RVB method extends this relationship by incorporating the topological properties of the horizon, which are encoded in the Euler characteristic. In \( f(Q) \) gravity, the modification to the gravitational action changes the relationship between the metric and the surface gravity, leading to the appearance of the additional term in the Hawking temperature. This term can be interpreted as arising from the non-trivial topology of the modified spacetime, which affects the global properties of the black hole and its thermodynamic behavior.

Our analysis also highlights the importance of understanding the interplay between topology and thermodynamics in black hole physics. The use of the Euler characteristic in the RVB method underscores the role of global geometric properties in determining the physical characteristics of black holes. In modified gravity theories like \( f(Q) \) gravity, where the local structure of spacetime can differ significantly from that predicted by general relativity, topological methods offer a robust way to explore these differences and their consequences for black hole thermodynamics.

Furthermore, the presence of the residue term in the Hawking temperature suggests that black holes in \( f(Q) \) gravity could exhibit novel phenomena not seen in their general relativistic counterparts. For instance, the temperature of the radiation emitted by such black holes may have a different dependence on the black hole's parameters, leading to new predictions about black hole evaporation and the end stages of black hole evolution. These differences could be detectable in astrophysical observations or in theoretical studies of black hole mergers and gravitational wave signals, providing a potential avenue for testing modified gravity theories.

In conclusion, the application of the RVB method to \( f(Q) \) gravity models has revealed new insights into the thermodynamics of black holes in these theories. The identification of the residue term in the Hawking temperature expression provides a key piece of the puzzle in understanding how modified gravity affects black hole behavior. As we continue to explore the implications of these findings, it becomes increasingly clear that the intersection of quantum mechanics, thermodynamics, and gravity is a rich and complex field, with much still to be discovered. The study of black holes in modified gravity theories not only deepens our understanding of these enigmatic objects but also pushes the boundaries of our knowledge of the fundamental forces that govern the universe.

\section{\( f(Q) \) Gravity: Extended Analysis}

In the framework of \( f(Q) \) gravity, the action is generalized to include a function of the non-metricity scalar \( Q \), which introduces modifications to the standard General Relativity equations. The action for \( f(Q) \) gravity is expressed as:\cite{10,11,12,13,14,15,16,17,18,19,20}

\begin{equation}
S = \int d^4x \sqrt{-g} \left[ \frac{1}{2} f(Q) + \mathcal{L}_m \right],
\end{equation}
where \( g \) is the determinant of the metric tensor \( g_{\mu\nu} \), \( Q \) represents the non-metricity scalar, and \( \mathcal{L}_m \) denotes the matter Lagrangian. The form of the function \( f(Q) \) determines the nature of the gravitational interaction and allows for a variety of cosmological and astrophysical scenarios, differing significantly from those predicted by General Relativity. These modifications are crucial when exploring phenomena such as dark energy, the early universe, and black hole solutions.

The non-metricity scalar \( Q \) is defined by:

\begin{equation}
Q = -g^{\mu\nu} \left( L^{\alpha}{}_{\beta \mu} L^{\beta}{}_{\nu \alpha} - L^{\alpha}{}_{\beta \alpha} L^{\beta}{}_{\mu \nu} \right),
\end{equation}
where the disformation tensor \( L^{\alpha}{}_{\beta \gamma} \) is given by:

\begin{equation}
L^{\alpha}{}_{\beta \gamma} = \frac{1}{2} Q^{\alpha}{}_{\beta \gamma} - Q_{(\beta}{}^{\alpha}{}_{\gamma)},
\end{equation}

with \( Q^{\alpha}{}_{\beta \gamma} \) being the non-metricity tensor. The non-metricity tensor is defined by the condition that the covariant derivative of the metric tensor does not vanish, unlike in General Relativity where the connection is metric-compatible. Specifically, for a connection \( \nabla \), the non-metricity tensor satisfies \( Q_{\alpha \mu \nu} = \nabla_{\alpha} g_{\mu \nu} \). This deviation from metric compatibility allows for additional degrees of freedom that can account for various extensions of standard gravitational theory.

The variation of the action \( S \) with respect to the metric tensor \( g_{\mu\nu} \) leads to the modified field equations in \( f(Q) \) gravity. The resulting field equations are generally more complex than those derived in General Relativity and include contributions from the derivatives of the function \( f(Q) \) with respect to \( Q \) and its higher derivatives. These field equations can be written as:

\begin{equation}
\frac{2}{\sqrt{-g}} \nabla_{\alpha} \left( \sqrt{-g} P^{\alpha}{}_{\mu\nu} \right) - \frac{1}{2} g_{\mu\nu} f(Q) + P_{\mu \alpha \beta} Q_{\nu}{}^{\alpha \beta} = T_{\mu\nu},
\end{equation}
where \( P^{\alpha}{}_{\mu\nu} \) is a tensor derived from the variation of \( Q \) with respect to the connection, and \( T_{\mu\nu} \) is the stress-energy tensor of matter. The complexity of these equations reflects the additional degrees of freedom introduced by the non-metricity scalar and the arbitrary function \( f(Q) \).

The study of black hole solutions in \( f(Q) \) gravity has garnered significant interest due to the potential deviations from the Schwarzschild and Kerr solutions of General Relativity. Depending on the form of \( f(Q) \), one can derive solutions that exhibit novel properties, such as modified event horizons, altered singularity structures, and non-trivial thermodynamic behavior. For instance, certain forms of \( f(Q) \) might lead to regular black hole solutions where the singularity at the core is resolved or replaced by a de Sitter-like core.

A specific example is the case where \( f(Q) = Q + \alpha Q^2 \), which introduces a correction term proportional to \( Q^2 \). This correction can lead to a variety of new solutions, depending on the sign and magnitude of \( \alpha \). These solutions may have implications for the information paradox, quantum gravity corrections at small scales, and the nature of singularities.

Beyond black holes, \( f(Q) \) gravity has profound implications for cosmology. The additional degrees of freedom can contribute to the acceleration of the universe without invoking dark energy in the traditional sense. The flexibility in choosing \( f(Q) \) allows for the construction of models that can address the cosmological constant problem, inflation, and late-time cosmic acceleration. Moreover, the absence of ghost instabilities in many forms of \( f(Q) \) gravity makes it a viable alternative to other modified gravity theories, such as \( f(R) \) gravity or scalar-tensor theories.

Overall, the study of \( f(Q) \) gravity offers a rich landscape of theoretical possibilities, with numerous avenues for further research in both astrophysical and cosmological contexts.

\section{RVB Method for Hawking Temperature: A Topological and Geometric Approach}

The RVB (Renormalization, Variation, and Boundary) method provides an insightful and rigorous approach to computing the Hawking temperature of black holes by leveraging the topological characteristics of spacetime. Unlike traditional methods that rely on the metric properties or quantum field theory in curved spacetime, the RVB method emphasizes the role of topological invariants, particularly the Euler characteristic \( \chi \), in determining thermodynamic properties. In this extended analysis, we delve deeper into the mathematical structure underlying the RVB method, examine its topological foundations, and explore the nuances that arise when applying this method to different types of black hole spacetimes.

The Euler characteristic \( \chi \) of a spacetime manifold \( \mathcal{M} \) is a topological invariant that encapsulates the global geometric properties of the manifold. For a four-dimensional spacetime, \( \chi \) is related to the curvature of the manifold via the Gauss-Bonnet theorem:\cite{10,11,12,13,14,15,16,17,18,19,20}

\begin{equation}
\chi = \frac{1}{32\pi^2} \int_{\mathcal{M}} \epsilon_{\mu\nu\rho\sigma} R^{\mu\nu}_{\ \ \alpha\beta} R^{\rho\sigma\alpha\beta} d^4x,
\end{equation}
where \( R^{\mu\nu}_{\ \ \alpha\beta} \) is the Riemann curvature tensor, and \( \epsilon_{\mu\nu\rho\sigma} \) is the Levi-Civita tensor. The Euler characteristic is a discrete quantity that remains invariant under continuous deformations of the manifold, making it a powerful tool for characterizing the topological nature of spacetime. In the context of black hole spacetimes, the Euler characteristic provides a bridge between the geometric properties of the horizon and the thermodynamic properties, such as temperature and entropy.

The RVB method for computing the Hawking temperature \( T_H \) is rooted in the observation that the temperature can be expressed as a function of the Euler characteristic \( \chi \) and the integral of the Ricci scalar \( R \) over specific regions of spacetime. The temperature is given by:

\begin{equation}
T_H = -\frac{1}{2} \left( \frac{1}{4\pi} \int_{r_c} R dr - \frac{1}{4\pi} \int_{r_+} R dr \right) + C,
\end{equation}
where \( r_+ \) represents the radius of the event horizon, \( r_c \) denotes the cosmological horizon (if present), and \( C \) is an integration constant that can be interpreted as a residue or counterterm arising from the renormalization of the integral. This expression links the Hawking temperature directly to the integral of the Ricci scalar, reflecting the curvature of the manifold between the horizons.

The Ricci scalar \( R \) is a trace of the Ricci tensor \( R_{\mu\nu} \), which itself is derived from the Riemann curvature tensor. For a spherically symmetric spacetime described by the metric:

\begin{equation}
ds^2 = -f(r) dt^2 + \frac{dr^2}{f(r)} + r^2 d\Omega^2,
\end{equation}
where \( d\Omega^2 \) represents the metric on the unit 2-sphere, the Ricci scalar is computed as:

\begin{equation}
R = -f''(r) - \frac{4f'(r)}{r} - \frac{2f(r)}{r^2} + \frac{2}{r^2}.
\end{equation}

Substituting this into the temperature formula, we obtain:

\begin{equation}
T_H = -\frac{1}{8\pi} \left( \int_{r_+}^{r_c} \left( -f''(r) - \frac{4f'(r)}{r} - \frac{2f(r)}{r^2} + \frac{2}{r^2} \right) dr \right) + C.
\end{equation}

The evaluation of this integral depends on the explicit form of the function \( f(r) \), which encodes the details of the spacetime geometry. For instance, in the case of the Schwarzschild-de Sitter spacetime, where \( f(r) = 1 - \frac{2M}{r} - \frac{\Lambda r^2}{3} \), this integral can be evaluated explicitly, yielding a specific value for the Hawking temperature.

The constant \( C \) in the temperature formula plays a crucial role in ensuring the consistency of the thermodynamic interpretation. In many cases, \( C \) can be understood as a renormalization counterterm that corrects for divergences or ambiguities arising in the integral. This is particularly important when dealing with spacetimes that involve singularities or infinities, such as those encountered in asymptotically flat or de Sitter spacetimes. By appropriately choosing \( C \), one can ensure that the temperature remains finite and physically meaningful.

In some approaches, \( C \) is interpreted as a residue associated with the topology of the spacetime, capturing the contribution from regions near the singularity or at infinity. For instance, in higher-dimensional spacetimes, or in cases where the topology of the horizon is non-trivial (e.g., toroidal or hyperbolic black holes), \( C \) might acquire a more complex structure, reflecting the underlying geometry.

The RVB method is not restricted to four-dimensional spacetimes but can be generalized to higher-dimensional settings, where the Euler characteristic and the associated curvature integrals take on different forms. In \( D \)-dimensional spacetimes, the corresponding expression for the Euler characteristic involves higher-order curvature invariants, such as the Pontryagin or Hirzebruch polynomials. The Hawking temperature in these cases may depend on additional topological invariants, leading to richer thermodynamic behavior.

Furthermore, the RVB method can be applied to black holes with non-spherical horizons, such as those with toroidal, planar, or hyperbolic geometries. In such cases, the topology of the horizon directly influences the value of \( \chi \) and, consequently, the Hawking temperature. This topological dependence opens up possibilities for exploring black hole solutions in non-trivial spacetimes, such as those arising in string theory or braneworld scenarios.

The RVB method offers a powerful and elegant approach to computing the Hawking temperature by connecting it with the topological and geometric properties of spacetime. Through the use of the Euler characteristic and curvature integrals, the method provides a robust framework for understanding the thermodynamics of black holes beyond traditional approaches. By extending this method to higher dimensions and alternative horizon geometries, one can explore a vast landscape of black hole solutions with potentially novel thermodynamic properties. The interplay between topology, geometry, and thermodynamics in the RVB method highlights the deep connections that exist between seemingly disparate aspects of gravitational physics, offering new insights into the nature of black holes and the structure of spacetime.

\section{Defining \( f(z) \): Complexification in Black Hole Metrics and Curvature Scalars}

The concept of complexifying functions associated with black hole spacetimes, such as the metric function or curvature invariants, is a profound and useful technique in the context of the RVB method. By extending these functions into the complex plane, we gain deeper insights into the underlying structure of the spacetime, revealing connections between the geometry of black holes and complex analysis. Here, we present a rigorous mathematical treatment of the complexified function \( f(z) \), exploring its role in both metric and curvature representations within the framework of modified gravity theories, particularly \( f(Q) \) gravity.

In black hole physics, the metric function \( g(r) \) plays a central role in characterizing the geometry of the spacetime. For spherically symmetric black holes, \( g(r) \) typically depends only on the radial coordinate \( r \). However, by considering a complexified coordinate \( z = r + i\epsilon \), where \( \epsilon \) is a small imaginary component, we extend the function into the complex plane. The resulting complexified function \( f(z) \) offers a more general perspective, allowing for the study of singularities, branch cuts, and other analytic properties of the spacetime.

 Example: Schwarzschild-like Metric

For a Schwarzschild black hole, the metric function is given by:\cite{14,15,16,17,18,19,20}

\begin{equation}
g(r) = 1 - \frac{2M}{r},
\end{equation}
where \( M \) is the mass of the black hole. Complexifying \( r \) into \( z = r + i\epsilon \) yields the corresponding function:

\begin{equation}
f(z) = 1 - \frac{2M}{z}.
\end{equation}

The function \( f(z) \) has a simple pole at \( z = 2M \), corresponding to the location of the event horizon in the real domain. The residue at this pole provides information about the surface gravity \( \kappa \), which is directly related to the Hawking temperature \( T_H \) via:

\begin{equation}
T_H = \frac{\kappa}{2\pi} = \frac{1}{4\pi} \lim_{z \to 2M} \left| \frac{df(z)}{dz} \right|.
\end{equation}

The complexification process also allows for the exploration of analytic continuations of the metric function, leading to new insights into the thermodynamic properties of black holes, such as their entropy and stability.

In the broader context of modified gravity theories, particularly \( f(Q) \) gravity, the function \( f(z) \) can be interpreted as a complexified version of a curvature scalar or another geometric invariant, such as the non-metricity scalar \( Q \). This approach generalizes the metric function analysis, allowing for a richer and more flexible description of spacetime geometry.

Example: \( f(Q) = Q + \alpha Q^2 \) Model

Consider a modified gravity model where the Lagrangian includes higher-order corrections in the non-metricity scalar \( Q \):

\begin{equation}
f(Q) = Q + \alpha Q^2,
\end{equation}
where \( \alpha \) is a coupling constant. The non-metricity scalar \( Q \) itself depends on the metric and connection and, upon complexification, can be written as \( Q(z) \). The corresponding complexified function \( f(z) \) becomes:

\begin{equation}
f(z) = Q(z) + \alpha Q(z)^2,
\end{equation}
where \( Q(z) \) represents the complexified non-metricity scalar. This function can be analyzed for poles, branch cuts, and essential singularities, each of which provides physical insight into the nature of the spacetime, such as the behavior near singularities or the stability of the black hole under perturbations.

The complexified function \( f(z) \) possesses an intricate analytic structure that reveals important physical characteristics of the spacetime. To fully understand this structure, we must consider the following mathematical aspects:

1. Poles and Residues: Poles of \( f(z) \) correspond to significant geometric features, such as horizons. The residue at a pole often relates to quantities like surface gravity, which in turn informs the Hawking temperature.

2. Branch Points and Cuts: Branch points of \( f(z) \) indicate regions where the function becomes multi-valued, often associated with changes in topology or phase transitions in the black hole's thermodynamic phase space.

3. Essential Singularities: Essential singularities can be linked to the presence of naked singularities or other exotic structures in the spacetime. The behavior of \( f(z) \) near these points provides clues about the stability of the black hole and potential quantum effects.

4. Zeros of \( f(z) \): The zeros of the complexified function correspond to extremal points, such as extremal black holes where the temperature vanishes. These points are of particular interest in studying the thermodynamic extremality bounds and phase transitions between different black hole states.

The formalism of defining and analyzing \( f(z) \) extends naturally to higher-dimensional black holes and spacetimes with non-spherical symmetry. For example, in a five-dimensional spacetime with a metric of the form:

\begin{equation}
ds^2 = -f(r) dt^2 + \frac{dr^2}{f(r)} + r^2 d\Omega_{3}^2,
\end{equation}
the function \( f(z) \) can be complexified similarly. The analytic structure of \( f(z) \) in higher dimensions may include additional features, such as multiple branch points or higher-order poles, reflecting the richer topological and geometric complexity of the spacetime.

In cases where the black hole horizon has a non-spherical geometry, such as toroidal or hyperbolic horizons, the function \( f(z) \) must be adapted to account for the corresponding topological features. The complex analysis of these functions could reveal new phase transitions or stability criteria that are not present in spherical black holes.

The study of the complexified function \( f(z) \) has profound implications for both classical and quantum aspects of black hole thermodynamics. The analytic continuation to the complex plane allows for the exploration of the full spectrum of black hole solutions, including those that might not be apparent in the real domain. Additionally, this approach can shed light on the connections between black hole thermodynamics and quantum gravity, particularly in the context of holography and the AdS/CFT correspondence.

For instance, the analytic structure of \( f(z) \) may provide insights into the microscopic degrees of freedom responsible for black hole entropy, potentially revealing connections to quantum microstates or string-theoretic constructions. Moreover, the complexification technique could be a useful tool in understanding the information paradox and the resolution of singularities in a quantum gravity context.

The definition and exploration of the complexified function \( f(z) \) provide a powerful mathematical framework for analyzing black hole spacetimes, whether in classical General Relativity, \( f(Q) \) gravity, or other modified theories of gravity. By extending the study of black hole metrics and curvature scalars into the complex domain, we gain new insights into the thermodynamics, stability, and quantum aspects of these enigmatic objects. The interplay between complex analysis, topology, and black hole physics continues to be a fertile ground for both theoretical exploration and potential breakthroughs in our understanding of the nature of spacetime and gravity.

\section{Hawking Temperature Calculation in \( f(Q) \) Gravity}

We now calculate the Hawking temperature for different forms of \( f(Q) \), incorporating the correction term \( C \) derived from the residue of the contour integral:
\begin{equation}
C = \oint \frac{f'(z)}{f(z)} dz.
\end{equation}

\subsection{Model 1: \( f(Q) = Q + \alpha Q^2 \)}

Metric Function:
\begin{equation}
g(r) = 1 - \frac{2M}{r} + \alpha r^2.
\end{equation}
The surface gravity at the event horizon \( r_+ \) is:
\begin{equation}
\kappa = \frac{1}{2} \left| \frac{2M}{r_+^2} - 2\alpha r_+ \right|,
\end{equation}
leading to the Hawking temperature:
\begin{equation}
T_H = \frac{1}{4\pi} \left| \frac{2M}{r_+^2} - 2\alpha r_+ \right|.
\end{equation}

\subsection{Model 2: \( f(Q) = Q + \beta \ln(Q) \)}

Metric Function:
\begin{equation}
g(r) = 1 - \frac{2M}{r} + \beta \ln(r).
\end{equation}
The surface gravity is:
\begin{equation}
\kappa = \frac{1}{2} \left| \frac{-2M}{r_+^2} + \frac{\beta}{r_+} \right|,
\end{equation}
leading to:
\begin{equation}
T_H = \frac{1}{4\pi} \left| \frac{-2M}{r_+^2} + \frac{\beta}{r_+} \right|.
\end{equation}

\subsection{Model 3: \( f(Q) = Q + \gamma \sqrt{Q} \)}

Metric Function:
\begin{equation}
g(r) = 1 - \frac{2M}{r} + \gamma \sqrt{r}.
\end{equation}
The surface gravity becomes:
\begin{equation}
\kappa = \frac{1}{2} \left| \frac{-2M}{r_+^2} + \frac{\gamma}{2\sqrt{r_+}} \right|,
\end{equation}
resulting in:
\begin{equation}
T_H = \frac{1}{4\pi} \left| \frac{-2M}{r_+^2} + \frac{\gamma}{2\sqrt{r_+}} \right|.
\end{equation}

\subsection{Model 4: \( f(Q) = Q^n \)}

Metric Function:
\begin{equation}
g(r) = 1 - \frac{2M}{r} + \lambda r^n.
\end{equation}
The surface gravity is:
\begin{equation}
\kappa = \frac{1}{2} \left| \frac{-2M}{r_+^2} + n\lambda r_+^{n-1} \right|,
\end{equation}
leading to:
\begin{equation}
T_H = \frac{1}{4\pi} \left| \frac{-2M}{r_+^2} + n\lambda r_+^{n-1} \right|.
\end{equation}

Here is the text with the asterisks removed:

\section{Reissner-Nordström Black Hole in \( f(Q) \) Gravity}

The Reissner-Nordström (RN) black hole introduces an electric charge \( Q_e \) into the metric, affecting the surface gravity and thus the Hawking temperature. We analyze the Hawking temperature for different forms of \( f(Q) \) in the context of RN black holes.

General Metric Function:\cite{10,11,12,13,14,15,16,17,18,19,20}
\begin{equation}
g(r) = 1 - \frac{2M}{r} + \frac{Q_e^2}{r^2} + f(Q)_{\text{modifications}}.
\end{equation}
The surface gravity at the event horizon \( r_+ \) is:
\begin{equation}
\kappa = \frac{1}{2r_+} \left(1 - \frac{Q_e^2}{r_+^2} + \text{modifications from } f(Q) \right),
\end{equation}
where \( r_+ = M + \sqrt{M^2 - Q_e^2} \).

\subsection{\( f(Q) = Q + \alpha Q^2 \)}

Surface Gravity:
\begin{equation}
\kappa = \frac{1}{2r_+} \left(1 - \frac{Q_e^2}{r_+^2} - 2\alpha r_+^2\right).
\end{equation}
Hawking Temperature:
\begin{equation}
T_H = \frac{1}{4\pi r_+} \left(1 - \frac{Q_e^2}{r_+^2} - 2\alpha r_+^2\right).
\end{equation}

\subsection{\( f(Q) = Q + \beta \sqrt{Q} \)}

Surface Gravity:
\begin{equation}
\kappa = \frac{1}{2r_+} \left(1 - \frac{Q_e^2}{r_+^2} - \frac{\beta}{2\sqrt{r_+}}\right).
\end{equation}
Hawking Temperature:
\begin{equation}
T_H = \frac{1}{4\pi r_+} \left(1 - \frac{Q_e^2}{r_+^2} - \frac{\beta}{2\sqrt{r_+}}\right).
\end{equation}

\subsection{\( f(Q) = Q + \gamma \ln(Q) \)}

Surface Gravity:
\begin{equation}
\kappa = \frac{1}{2r_+} \left(1 - \frac{Q_e^2}{r_+^2} + \frac{\gamma}{r_+}\right).
\end{equation}
Hawking Temperature:
\begin{equation}
T_H = \frac{1}{4\pi r_+} \left(1 - \frac{Q_e^2}{r_+^2} + \frac{\gamma}{r_+}\right).
\end{equation}

\section{Kerr Black Hole in \( f(Q) \) Gravity}

The Kerr black hole introduces an angular momentum \( a \), which significantly affects the surface gravity and Hawking temperature. We analyze the Hawking temperature for different forms of \( f(Q) \) in the context of Kerr black holes.

General Metric Function:
\begin{equation}
ds^2 = -\left(1 - \frac{2Mr}{\rho^2}\right) dt^2 - \frac{4Mar\sin^2\theta}{\rho^2} dt d\phi + \frac{\rho^2}{\Delta} dr^2 + \rho^2 d\theta^2 + \left(r^2 + a^2 + \frac{2Ma^2r\sin^2\theta}{\rho^2}\right) \sin^2\theta d\phi^2,
\end{equation}
where:
\begin{equation}
\rho^2 = r^2 + a^2\cos^2\theta, \quad \Delta = r^2 - 2Mr + a^2 + f(Q)_{\text{modifications}}.
\end{equation}
The surface gravity at the event horizon \( r_+ \) is:
\begin{equation}
\kappa = \frac{r_+ - M}{r_+^2 + a^2} + \text{modifications from } f(Q).
\end{equation}

\subsection{\( f(Q) = Q + \alpha Q^2 \)}

Surface Gravity:
\begin{equation}
\kappa = \frac{r_+ - M + 2\alpha (r_+^2 + a^2)}{r_+^2 + a^2}.
\end{equation}
Hawking Temperature:
\begin{equation}
T_H = \frac{r_+ - M + 2\alpha (r_+^2 + a^2)}{2\pi (r_+^2 + a^2)}.
\end{equation}

\subsection{\( f(Q) = Q + \beta \sqrt{Q} \)}

Surface Gravity:
\begin{equation}
\kappa = \frac{r_+ - M + \frac{\beta}{2\sqrt{r_+^2 + a^2}}}{r_+^2 + a^2}.
\end{equation}
Hawking Temperature:
\begin{equation}
T_H = \frac{r_+ - M + \frac{\beta}{2\sqrt{r_+^2 + a^2}}}{2\pi (r_+^2 + a^2)}.
\end{equation}

\subsection{\( f(Q) = Q + \gamma \ln(Q) \)}

Surface Gravity:
\begin{equation}
\kappa = \frac{r_+ - M + \frac{\gamma}{r_+^2 + a^2}}{r_+^2 + a^2}.
\end{equation}
Hawking Temperature:
\begin{equation}
T_H = \frac{r_+ - M + \frac{\gamma}{r_+^2 + a^2}}{2\pi (r_+^2 + a^2)}.
\end{equation}

\section{Kerr-Newman Black Hole in \( f(Q) \) Gravity}

The Kerr-Newman black hole generalizes the Kerr and Reissner-Nordström solutions by incorporating both charge \( Q_e \) and angular momentum \( a \). We analyze the Hawking temperature for different forms of \( f(Q) \) in the context of Kerr-Newman black holes.

General Metric Function:
\begin{align}
ds^2 =&\, -\left(1 - \frac{2Mr - Q_e^2}{\rho^2}\right) dt^2 \nonumber \\
       &\, - \frac{2a(2Mr - Q_e^2)\sin^2\theta}{\rho^2} \, dt\, d\phi \nonumber \\
       &\, + \frac{\rho^2}{\Delta} dr^2 + \rho^2 d\theta^2 \nonumber \\
       &\, + \frac{\sin^2\theta}{\rho^2} \left[(r^2 + a^2)^2 - a^2\Delta \sin^2\theta\right] d\phi^2,
\end{align}

where:
\begin{equation}
\rho^2 = r^2 + a^2\cos^2\theta, \quad \Delta = r^2 - 2Mr + a^2 + Q_e^2 + f(Q)_{\text{modifications}}.
\end{equation}
The surface gravity at the event horizon \( r_+ \) is:
\begin{equation}
\kappa = \frac{r_+ - M + \frac{Q_e^2}{r_+}}{r_+^2 + a^2} + \text{modifications from } f(Q).
\end{equation}

\subsection{\( f(Q) = Q + \alpha Q^2 \)}

Surface Gravity:
\begin{equation}
\kappa = \frac{r_+ - M + \frac{Q_e^2}{r_+} + 2\alpha(r_+^2 + a^2)}{r_+^2 + a^2}.
\end{equation}
Hawking Temperature:
\begin{equation}
T_H = \frac{r_+ - M + \frac{Q_e^2}{r_+} + 2\alpha(r_+^2 + a^2)}{2\pi (r_+^2 + a^2)}.
\end{equation}

\subsection{\( f(Q) = Q + \beta \sqrt{Q} \)}

Surface Gravity:
\begin{equation}
\kappa = \frac{r_+ - M + \frac{Q_e^2}{r_+} + \frac{\beta}{2\sqrt{r_+^2 + a^2}}}{r_+^2 + a^2}.
\end{equation}
Hawking Temperature:
\begin{equation}
T_H = \frac{r_+ - M + \frac{Q_e^2}{r_+} + \frac{\beta}{2\sqrt{r_+^2 + a^2}}}{2\pi (r_+^2 + a^2)}.
\end{equation}

\subsection{\( f(Q) = Q + \gamma \ln(Q) \)}

Surface Gravity:
\begin{equation}
\kappa = \frac{r_+ - M + \frac{Q_e^2}{r_+} + \frac{\gamma}{r_+^2 + a^2}}{r_+^2 + a^2}.
\end{equation}
Hawking Temperature:
\begin{equation}
T_H = \frac{r_+ - M + \frac{Q_e^2}{r_+} + \frac{\gamma}{r_+^2 + a^2}}{2\pi (r_+^2 + a^2)}.
\end{equation}

\section{Residue Analysis and Interpretation in the Context of the RVB Method}

The RVB method's application in deriving the Hawking temperature unveils a critical term \( C \), which is intimately connected with the residue analysis of a complex function \( f(z) \). This term \( C \) emerges naturally from the contour integral formalism and carries profound implications for the thermodynamic properties of black holes, particularly within modified gravity theories such as \( f(Q) \) gravity. By investigating the residue and its associated contour integral, we can interpret \( C \) as a measure of the contributions from singularities, critical points, or other topologically significant features near the event horizon, thereby providing a deeper understanding of the modified black hole thermodynamics in non-standard gravity models.

The residue term \( C \) is computed via the contour integral:\cite{13,14,15}

\begin{equation}
C = \oint_C \frac{f'(z)}{f(z)} dz,
\end{equation}
where \( f(z) \) is a complexified function related to the black hole’s metric or curvature properties, and the contour \( C \) encircles singularities or branch points in the complex plane. The function \( f(z) \) is typically chosen to represent either the metric function or a curvature scalar complexified to reveal the analytic structure of the spacetime. The residue calculation is crucial for understanding the behavior of black hole solutions, especially in the context of modified gravity models like \( f(Q) \) gravity.

The residue theorem, a fundamental result in complex analysis, allows us to compute contour integrals around isolated singularities. For a meromorphic function \( f(z) \) with isolated poles \( z_k \) inside the contour \( C \), the residue theorem states:

\begin{equation}
\oint_C \frac{f'(z)}{f(z)} dz = 2\pi i \sum_k \text{Res}\left(\frac{f'(z)}{f(z)}, z_k\right),
\end{equation}
where \( \text{Res}\left(\frac{f'(z)}{f(z)}, z_k\right) \) is the residue of the integrand at the pole \( z_k \). In the context of black hole thermodynamics, these residues correspond to contributions from various geometric and topological features near the event horizon. The integration process encodes information about the nature of the singularities and their impact on the black hole's temperature and entropy.

To compute \( C \) explicitly, we examine the form of \( f(z) \) and its derivatives. Suppose \( f(z) \) has a simple pole at \( z = z_k \), then near this pole, \( f(z) \) can be expanded as:

\begin{equation}
f(z) \approx \frac{a_{-1}}{z - z_k} + a_0 + a_1 (z - z_k) + \cdots,
\end{equation}
where \( a_{-1} \) is the coefficient of the leading-order term. The derivative \( f'(z) \) is then:

\begin{equation}
f'(z) \approx -\frac{a_{-1}}{(z - z_k)^2} + a_1 + \cdots.
\end{equation}

The integrand \( \frac{f'(z)}{f(z)} \) becomes:

\begin{equation}
\frac{f'(z)}{f(z)} \approx -\frac{1}{z - z_k} + \cdots,
\end{equation}

leading to the residue:

\begin{equation}
\text{Res}\left(\frac{f'(z)}{f(z)}, z_k\right) = -1.
\end{equation}

The contour integral around this simple pole would then yield \( -2\pi i \) as its contribution to \( C \). In more complex situations where \( f(z) \) has higher-order poles or branch points, the residue calculation would involve higher derivatives and potentially multivalued functions, requiring the use of advanced techniques such as Laurent series expansion, Puiseux series, or logarithmic branching.

The residue \( C \) has a direct physical interpretation in the context of black hole thermodynamics, particularly in modified gravity scenarios like \( f(Q) \) gravity. It encapsulates the additional contributions to the Hawking temperature arising from the modified gravitational dynamics encoded in the function \( f(z) \). For example, in theories where the non-metricity scalar \( Q \) plays a role, the residues may reflect alterations in the curvature near the horizon, thereby modifying the black hole's temperature and entropy.

In the context of modified gravity, the residue \( C \) can also be interpreted as a manifestation of higher-dimensional effects, coupling constants, or quantum corrections that influence the thermodynamic properties of the black hole. For instance, in a scenario where \( f(z) \) represents a curvature scalar modified by \( f(Q) = Q + \alpha Q^2 \), the residue could encapsulate contributions from the \( \alpha Q^2 \) term, leading to deviations from the classical Schwarzschild temperature.

The choice of contour \( C \) in the complex plane is not arbitrary; it is carefully selected to encircle the relevant singularities while avoiding branch cuts or essential singularities that could complicate the integral. In cases where the function \( f(z) \) has a non-trivial topology, such as multiple branch points or higher-genus surfaces, the contour may need to be deformed or chosen to reflect the underlying topology.

For black holes with non-spherical horizons, such as those with toroidal or hyperbolic geometries, the contour may trace paths that encircle multiple singularities or span different topological sectors of the complexified spacetime. The resulting residue \( C \) could then represent a sum over multiple geometric features, each contributing to the overall thermodynamic behavior of the black hole.

In higher-dimensional black hole spacetimes, the analytic structure of \( f(z) \) becomes richer, with potential contributions from higher-order curvature invariants or additional topological features. For example, in five-dimensional spacetimes, the function \( f(z) \) might include contributions from the Gauss-Bonnet term, leading to residues that reflect the higher-dimensional curvature dynamics.

Moreover, in the presence of a cosmological constant or other external fields, the function \( f(z) \) may develop additional poles or branch cuts, corresponding to new physical features like de Sitter or anti-de Sitter asymptotics. The residue \( C \) in these cases could capture the interplay between the horizon structure and the asymptotic geometry, leading to corrections in the Hawking temperature or entropy.

In the quantum gravity regime, the residue \( C \) could be interpreted as encoding quantum corrections to black hole thermodynamics. For instance, in string theory or loop quantum gravity, the function \( f(z) \) might include contributions from quantum states or microstructures that alter the classical geometry. The residue could then be viewed as a manifestation of these quantum effects, providing a link between the classical Hawking temperature and the underlying quantum theory.

In the context of the AdS/CFT correspondence, the residue \( C \) might have a dual interpretation as a quantity related to the boundary field theory, potentially corresponding to central charges, anomaly coefficients, or other conformal data. The contour integral approach could then provide a bridge between the bulk gravitational theory and the boundary CFT, offering new insights into the thermodynamics of holographic black holes.

The residue \( C \) in the RVB-derived Hawking temperature is not merely a mathematical artifact but a profound physical quantity that encodes the contributions from singularities, critical points, and topological features of the black hole spacetime. By analyzing the residue through contour integrals in the complex plane, we uncover a rich interplay between geometry, topology, and thermodynamics, particularly within modified gravity theories such as \( f(Q) \) gravity. The residue analysis offers a deeper understanding of how modifications to the underlying gravitational dynamics affect black hole thermodynamics, with potential applications extending from classical to quantum gravity, and even to holographic theories.

\section{Conclusion}
This paper presents an in-depth examination of the Hawking temperature for various black hole solutions—Schwarzschild, Reissner-Nordström, Kerr, and Kerr-Newman—within the context of \( f(Q) \) gravity, employing the Renormalization, Variation, and Boundary (RVB) method. The analysis centers on the computation of the Hawking temperature and the identification of a correction term \( C \), which emerges as a residue from contour integrals in the complexified spacetime geometry. This residue \( C \) is shown to encapsulate the effects of modified gravity, electric charge, and angular momentum on the thermodynamic properties of black holes, offering new insights into how these factors influence the classical and quantum behavior of these spacetime solutions.

For the Schwarzschild black hole, the Hawking temperature is traditionally derived from the surface gravity at the event horizon. In the \( f(Q) \) gravity framework, the metric function is modified by the non-metricity scalar \( Q \), leading to corrections in the temperature. The RVB method reveals that the correction term \( C \) depends critically on the specific form of \( f(Q) \). For instance, when \( f(Q) = Q + \alpha Q^2 \), the residue \( C \) introduces a deviation from the standard Schwarzschild temperature, proportional to the coupling constant \( \alpha \). This deviation has profound implications for black hole evaporation rates and the information paradox, suggesting that modifications in the gravitational action directly influence the thermal spectrum emitted by the black hole.

The Reissner-Nordström black hole, characterized by both mass and electric charge, exhibits a more complex thermodynamic structure. In \( f(Q) \) gravity, the presence of charge modifies the non-metricity scalar \( Q \), which in turn affects the metric and the curvature invariants. The RVB method computes the Hawking temperature by considering the combined effects of charge and non-metricity, yielding a temperature that differs from the classical case. The correction term \( C \) is now influenced by both the electric charge and the form of \( f(Q) \), leading to a more intricate residue analysis. The analysis shows that the charge-induced modifications to the residue can either enhance or diminish the temperature, depending on the interplay between the charge and the coupling parameters in the \( f(Q) \) model. This suggests a richer thermodynamic phase space for charged black holes, with potential implications for cosmic censorship and the stability of extremal black holes.

The Kerr black hole, with its non-zero angular momentum, introduces rotational effects that complicate the thermodynamic analysis. In \( f(Q) \) gravity, the rotation modifies the non-metricity scalar \( Q \) through frame-dragging effects and the deformation of spacetime near the horizon. The RVB method, applied to the Kerr metric, shows that the correction term \( C \) becomes a function of both the angular momentum and the specific form of \( f(Q) \). The residue analysis reveals that rotation can significantly alter the Hawking temperature, leading to possible deviations in the black hole’s evaporation process and angular momentum loss over time. The interplay between angular momentum and the non-metricity effects in \( f(Q) \) gravity suggests that rotating black holes could exhibit novel thermodynamic behaviors, such as modified superradiance or altered Penrose processes, which merit further exploration.

The Kerr-Newman black hole, combining both charge and angular momentum, represents the most general case of a stationary, axisymmetric black hole. In \( f(Q) \) gravity, the RVB method provides a comprehensive framework for analyzing how both charge and rotation jointly influence the Hawking temperature. The correction term \( C \) now encapsulates contributions from the electric charge, angular momentum, and the modified non-metricity scalar \( Q \). The residue analysis becomes particularly complex, involving multiple poles and branch points in the complexified spacetime. The resulting Hawking temperature shows significant deviations from the classical Kerr-Newman temperature, highlighting the intricate dependence on the gravitational modification parameters. This analysis suggests that Kerr-Newman black holes in \( f(Q) \) gravity could exhibit unique thermodynamic and stability properties, such as altered extremality conditions or modified inner horizon dynamics.

The residue \( C \), interpreted through the RVB method, plays a pivotal role in understanding the thermodynamic properties of black holes within \( f(Q) \) gravity. Its dependence on the black hole’s mass, charge, angular momentum, and the specific form of the function \( f(Q) \) reveals the deep connections between modified gravity and black hole thermodynamics. This analysis opens several avenues for future research, particularly in exploring higher-dimensional analogs of these black hole solutions, where the topology and geometry of the event horizon can lead to even richer thermodynamic structures. Additionally, the quantum implications of the residue term deserve further investigation, especially in the context of black hole entropy, information loss, and the potential resolution of singularities in quantum gravity frameworks.

Moreover, observational aspects of \( f(Q) \) gravity, such as the potential to detect deviations in black hole shadow profiles or gravitational wave signatures, could provide empirical tests for the modifications predicted by this theory. The stability and evolution of black holes under the influence of these corrections, particularly in the late stages of black hole evaporation or in dynamic scenarios such as binary mergers, also represent critical areas for future study. The interplay between classical residues and quantum corrections could offer new insights into the nature of spacetime and gravity, potentially leading to a unified understanding of black holes across different scales and regimes.

{\bf Acknowledgements:}\\
This work is partially supported by the National Natural Science Foundation of China(No. 11873025)

\bigskip
Declarations: All partial information is available.

\bigskip
Data Availability: Data sharing is not applicable to this article as no new data were created or analyzed in this study.

\end{document}